\documentclass[12pt,preprint]{aastex}

\shorttitle{Planet Around HD 17092.}
\shortauthors{Niedzielski et al.}

\begin{document}

\title{A Planetary Mass Companion to the K0 Giant HD 17092. }

\author{A. Niedzielski\altaffilmark{1,2},  M. Konacki\altaffilmark{3}, A. Wolszczan\altaffilmark{2,1}, G. Nowak\altaffilmark{1},  G. Maciejewski\altaffilmark{1}, C. R. Gelino\altaffilmark{5}, M. Shao\altaffilmark{6}, M. Shetrone\altaffilmark{4}, L. W. Ramsey\altaffilmark{2} }

\altaffiltext{1}{Toru\'n Centre for Astronomy, Nicolaus Copernicus University, ul. Gagarina  11, 87-100 Toru\'n, Poland}
\altaffiltext{2}{Department of Astronomy and Astrophysics, Pennsylvania State University, 525 Davey Laboratory, University Park, PA 16802.}
\altaffiltext{3}{Nicolaus Copernicus Astronomical Center, Rabianska 7,   87-100 Toru\'n, Poland}
\altaffiltext{4}{McDonald Observatory, University of Texas, Fort Davis, TX 79734}
\altaffiltext{5}{Spitzer Science Center, MC 220-6, California Institute of Technology, 1200 E. California Blvd, Pasadena, CA 91125}
\altaffiltext{6}{Jet Propulsion Laboratory, California Institute of Technology, 4800 Oak Grove Drive, Pasadena, CA 91109}

\begin{abstract}
We report the discovery of a substellar-mass companion to the K0-giant HD 17092
with the Hobby-Eberly Telescope. In the absence of any correlation of the observed
360-day periodicity with the standard
indicators of stellar activity, the observed radial velocity variations
are most plausibly explained in terms
of a Keplerian motion of a planetary-mass body around the star. With the estimated
stellar mass of 2.3M$_\odot$, the minimum mass of the planet is 4.6M$_J$.
The planet's orbit
is characterized by a mild eccentricity of $e$=0.17 and
a semi-major axis of 1.3 AU. This is the tenth published detection of
a planetary companion around a red giant star. Such discoveries add
to our understanding of planet formation around intermediate-mass stars and
they provide dynamical information on the evolution of planetary systems around
post-main sequence stars.
\end{abstract}

\keywords{planetary systems-stars: individual (HD 17092)}

\section{Introduction}

After a decade of searches for planets around Sun-like stars, it has become
apparent that to achieve a satisfying level of understanding of planet formation and evolution,
the surveys have to be extended to other types of stars. So far, the most successful of these 
have been searches for planets around low-mass dwarfs, which, among other goals, have been driven
by anticipation that Earth-mass planets can be found in their habitable zones
\citep{2005PThPS.158...24M, 2005PThPS.158...43M}.
Surveys of white dwarfs, which probe ancient planetary systems -
survivors of the evolution of their parent stars - exemplify an extension of
planet searches to the endpoint of stellar evolution \citep[e.g.][]{Kepler05}.
Finally, searches for neutron star planets can
provide information on planets around massive stars \citep{1993ApJ...419L..65T}
and on planet formation in extreme, post-supernova environments \citep{2003ApJ...591L.147K, 2006Natur.440..772W}.

Yet another, so far meagerly explored area of the extrasolar planetary research
involves searches for planets around giant stars.
More than a decade ago, precision radial velocity (RV) studies have established that
GK-giant stars exhibit RV variations ranging from days to many hundreds of days
\citep[e.g.][]{walker89, HC93, 1994ApJ...422..366H}. Enough observational
evidence has been accumulated to identify three distinct sources of this variability,
namely stellar pulsations, surface activity and a presence of substellar companions.
As Doppler searches for planets around main sequence (MS)
stars become inefficient for spectral types earlier than F6-F8, because of paucity
of spectral features and their rotational broadening, extending studies of planetary
system formation and evolution to stellar masses substantially larger than 1 M$_{\odot}$ is 
observationally difficult. A potentially very efficient, indirect way to remove this difficulty
is to conduct surveys of post-MS giants. These evolved stars have cool atmospheres and many
narrow spectral lines, which can be utilized in RV measurements to give
an adequate precision level ($<$10 m s$^{-1}$). Discoveries of planets around post-MS
giants, in numbers comparable to the current statistics of planets around MS-dwarfs
\citep[e.g.][]{2006ApJ...646..505B} , will most certainly provide the much needed 
information on planet formation around intermediate mass MS-progenitors ($\geq 1.5 M_{\odot}$)
and they will create an experimental basis with which to study 
dynamics of planetary systems orbiting evolving stars \citep[e.g.][]{1998Icar..134..303D}. 
Sufficiently large surveys of post-MS giants should furnish enough
planet detections to meaningfully address the problem of a long-term survival of
planetary systems around stars that are off the MS and
on their way to the final white dwarf stage.

In order to address the above issues, we have joined the existing surveys 
\citep[e.g.][and references therein]{2006A&A...457..335H, sato07} 
with our own long-term project to search for planets around 
evolved stars with the 9.2-m Hobby-Eberly Telescope
and its High Resolution Spectrograph. The sample of stars we have been monitoring
since early 2004 is composed of two groups, approximately equal in numbers. The first one
falls in the ``clump giant'' region of the HR-diagram \citep{jim98}, which contains stars of various
masses over a range of evolutionary stages. The second group comprises stars, which have
recently left the MS and are located $\sim$1.5 mag above it. Generally, all
our targets, a total of $>$900 GK-giants brighter than $\sim$11 mag, occupy the area in
the HR-diagram, which is approximately defined by the MS, the instability strip, and the coronal
dividing line.
If the frequency of occurence of planets
around MS-progenitors of GK-giants is similar to that of planets around solar-type stars,
our survey should detect 50-100 planets and planetary systems, which,
together with the detections from similar projects, will provide a firm basis for studies
of planetary system formation and evolution around $>1M_{\odot}$ stars.

In this paper, we describe our survey and the detection of
a planetary-mass companion to the K0-giant HD 17092. Details of the observing procedure
and survey strategy are given in Section 2, followed by a description of the basic
properties of HD 17092 in Section 3. The analysis of radial velocity and line bisector
measurements, including a discussion of the HIPPARCOS photometry of the star, is given
in Section 4. Finally, our results are discussed and summarized in Section 5.

\section{Observations}

Observations were made between 2004, January and 2007, March, with the Hobby-Eberly Telescope (HET) \citep{lwr98} equipped with the High Resolution Spectrograph (HRS) \citep{tull98} in the queue scheduled mode \citep{HetQ}. The spectrograph was used in the R=60,000 resolution mode with a gas cell ($I_2$) inserted into the optical path, and it
was fed with a 2 arcsec fiber. 
The observing scheme followed standard practices implemented in precision radial
velocity measurements with the iodine cell \citep{iodine}. 
The spectra consisted of 46 echelle orders recorded on the ``blue'' CCD chip (407.6-592 nm) and 24 orders on the ``red'' one (602-783.8 nm). The spectral data 
used for RV measurements were extracted from the 17 orders, which cover the 505 to 592 nm range of the $I_2$ cell spectrum.

Originally,
HD 17092 had been observed as part of the astrometric reference star selection program related to
a search for terrestrial-mass planets with the Space Interferometry Mission \citep{sim_epics, 2005AIPC..752...38N}.
The star was added to the list of candidates for substellar companions, when it became clear that
it exhibited RV variations, which disqualified it as a potential astrometric reference standard.

The observing strategy is illustrated in Fig. 1a. Measurements of a particular
target star begin with 2-3 exposures, typically 3-6 months apart, to check for any RV variability
exceeding a 30-50 m s$^{-1}$ threshold. If a significant variability is detected, the star is scheduled
for more frequent observations, and, if the RV variability is confirmed, it becomes part of
the high priority list.

We have collected radial velocity measurements of HD 17092 at 59 epochs.
Typically, the signal-to-noise ratio per resolution element in the spectra
was 200-250 at 594 nm in 3-8 minutes, depending on the atmospheric conditions. 
The basic data reduction was performed
 using standard IRAF\footnote{IRAF is distributed by the National Optical Astronomy Observatories, which are operated by the Association of Universities for Research in Astronomy, Inc., under cooperative agreement with the National Science Foundation.} scripts.
Radial velocities were measured by means of the commonly used $I_2$ cell calibration technique
\citep{Butler+1996}. 
A template spectrum was constructed from a high-resolution Fourier Transform Spectrometer (FTS) $I_2$ spectrum and a high signal-to-noise
stellar spectrum measured without the $I_2$ cell. 
Doppler shifts were derived from least-square fits of template spectra to
stellar spectra with the imprinted $I_2$ absorption lines.
The resultant radial velocity measurement for each epoch was derived as a mean value of
the independent determinations from the 17 usable echelle orders. The corresponding
uncertainties of these measurements were calculated assuming that errors obeyed
the Student's t-distribution and they typically fell in a 4-5 m s$^{-1}$ range at 1$\sigma$-level.  Radial velocities were referred to the Solar System barycenter using the \citet{1980A&AS...41....1S} algorithm.

\section{The star}

HD 17092 (BD+49 767) was classified as a K0-star by \citet{hd}. These authors also measured its photographic and photovisual magnitudes to be Ptg=8${^m}$.8 and Ptm=7${^m}$.8, respectively. The Tycho-2 catalogue \citep{tycho2} lists the values of B${_T}$=9${^m}$.374$\pm$0.019 and V${_T}$=7${^m}$.875$\pm$0.011 derived from the HIPPARCOS observations of the star. In the Tycho-1 catalogue \citep{tycho1},values of V=7${^m}$.73, B-V=1${^m}$.247$\pm$0.014 are also given. In the Tycho star mapper experiment, a trigonometric parallax of HD 17092 was measured as 9.2$\pm$5.5 mas.

A detailed inspection of our spectra in the range of 515-520 nm (comparison of intensities and widths of MgI b triplet lines at 516.7, 517.2 and 518.4 nm) suggests that the star is a giant. The absolute visual magnitude of HD 17092 is M${_V}$=1${^m}$.76, assuming (B-V)${_0}$=1.${^m}$00 after \citet{sk82}, the above Tycho parallax, and R=3.1. This makes the star almost one magnitude fainter compared to a typical M${_V}$(K0III)=0${^m}$.8 \citep{sk82}. In what follows, we will assume that
HD 17092 is a typical giant with M$_V$=0$^m$.8, as this value remains consistent with the Tycho
parallax uncertainty.

The atmospheric parameters of HD 17092 were recently estimated by \citet{pz06} through the analysis of 230 FeI and 11 FeII lines in the optical spectra. The results, T${_{eff}}$=4650K, log(g)=3.0, and [Fe/H]=0.18, are very close to rough estimates obtained from the calibrations by \citet{sk81}, which give logT${_{eff}}$=3.681, and log(g)=2.89. 
\citet{pz06} have also estimated the mass of HD 17092 as M/M$_{\odot}$=2.3$\pm$0.3 by comparing the star's position in the HR diagram with evolutionary tracks of \citet{girardi2000} given the above metalicity. In absence of a direct measurement of the radius of HD 17092, we assume that it is similar to the radii of stars with [Fe/H]$\geq$ -0.5 and adopt R/R${_\odot}$=10.1$\pm$4.2 after \cite{alonso2000}.

The projected rotational velocity of HD 17092, v sin$i\leq$1 km s$^{-1}$, was estimated using the cross-correlation method \citep{Benz+Mayor1984}. From this value and the adopted stellar radius we have obtained an estimate of the rotation period of HD 17092, P$_{rot}\approx$505 days. This value appears to be typical for K0 giants \citep{deMedeiros1996}.

\section{Data analysis}

\subsection{Modeling of the companion orbit}

Radial velocity variations of HD 17092 over a 3-year period are shown in Fig. 1a, together
with the best-fit model of a Keplerian orbit. The best-fit parameters of the orbit
and their Monte Carlo estimated uncertainties are listed in Table 1. The residuals shown
in Fig. 1b are characterized by the rms value of $\sim$16 m $s^{-1}$. 
If the observed RV variations are indeed caused by an orbiting companion, it moves in
a $\sim$360-day, moderately eccentric orbit, with a semi-major axis of 1.3 A.U., and has
a miniumum mass of m$_2$ sin $i$=4.6$M_{J}$, for the assumed stellar mass of 2.3 $M_{\odot}$. 
This mass indicates a planetary origin of the object over a range of possible values
of sin $i$ extending beyond the median inclination of $i=60^{\circ}$ for randomly oriented
orbits. The reduced
$\chi^2$=10 for the fit suggests a presence of additional variations in the data,
which have also been reported for other planet detections around giants \citep[e.g.][]{2002ApJ...576..478F} and are not
uncommon for this type of stars \citep{Setiawanetal2004, 2005A&A...437..743H}. 
Our work also confirms that red giants exhibit a RV scatter at an average level of $\sim$20 m s$^{-1}$, as the result of a stochastic
intrinsic stellar activity \citep{NW2007}. 
To account for these variations, we have adopted
a conservative error estimation procedure by adding in quadrature a constant 15 m s$^{-1}$ error
to the RV measurement uncertainties, before performing a least-squares fit of the orbit.
This approach resulted in parameter error estimates (Table 2), which realistically absorb
any leftover, unmodeled RV variations in the data. Their nature will be further discussed, 
when more data become available.

\subsection{Line bisector and curvature analysis}

Precision radial velocity measurements may be significantly affected by phenomena
that are not related to stellar reflex motion caused by the presence of an orbiting planet.
Changes in line shapes arising from motions in the stellar atmosphere, related to non-radial pulsations or inhomogeneous convection and/or spots combined with rotation, or distortions induced by light contamination by an unseen stellar companion, can mimic low-level radial velocity variations. Therefore, especially in the case of giant stars, it is important to verify whether the observed radial velocity variations are real, or if they are possibly generated by changes in the symmetry of spectral lines due to the above effects.

The basic tool to study the origin of RV variations derived from stellar spectra is the analysis of shapes of the spectral lines using the line bisector technique \citep{1983PASP...95..252G, 2005PASP..117..711G}. It has been extensively
used in the process of confirmation of a planetary origin of RV variations in solar-type stars \citep{Hat1997, Hat1998a, Hat1998b} and it has become a mandatory part of the analysis of RV data from giants \citep{Set2003, 2005A&A...437L..31S, 2005A&A...437..743H,  Reffertetal2006, sato07}.
To examine line profile variations in HD 17092 we have selected several lines of a moderate intensity, which were free of blends and were located close to the centers of echelle orders. Because the HRS spectra extend far beyond the range occupied by the $I_2$ lines, we were able to measure line profiles in the same spectra which were used for radial velocity determination.

In order to avoid any contamination by the $I_2$ spectrum,
we have selected lines in the wavelength range redwards of 660 nm, which also meant that we
could not use the standard bisector line of Fe I at 625.256 nm. Instead, we have used
three lines, Ni I 664.638, Ni I 676.784, and Cr I 663.003, which were analyzed in detail to search for any possible systematic effects. In fact, \citet{Dall+2006} has found that the Ni I 664.3683
nm line shows larger bisector variations than the Fe I 625.256 nm line. Also, the other two lines
adopted for this analysis show well defined bisectors. We have measured
two line parameters, the line bisector span and its curvature, under the assumption that
the mean bisector and velocity span of the three lines reflect the upper limit 
of a possible atmospheric activity of the star. Uncertainties in the derived values of
the bisector span and curvature were estimated as standard deviations from the mean.
As shown in Fig. 2, both parameters, when plotted as a function of 
orbital phase, are not correlated with radial velocities. 

\subsection{Photometric variabilty}

The HIPPARCOS star mapper (TYCHO) has made 124 photometric observations of HD 17092 in V${_T}$ and B${_T}$ between JD 2447915.85944 and 2449039.86213, about 12 years before the beginning of our survey. The measured scatter in V${_T}$ (defined as 0.5(V$_{T}$(85)-V$_{T}$(15)), see \citet{tycho1} for details) was 0.092 mag without any sign of systematic variability.  The observed scatter is consistent with the precision of Tycho photometry for a given stellar magnitude range.

We have performed a deeper search for any possible periodicities in the Tycho photometry of 
the star by computing
a Lomb-Scargle periodogram of these data.  No significant peaks were found in the spectrum above the false positive probability of 0.5 (Fig. 3). In particular, no excess fluctuation power is present at and around the 360-day period detected in the RV data. 

\section{Discussion and conclusions}

In this paper, we have presented a compelling evidence that the K0-giant, HD 17092, exhibits
a strictly periodic radial velocity variation with the period of 360$\pm$2 days over a 1200-day
span of observations. When interpreted in terms of a Keplerian motion, this periodicity indicates
the presence of a sub-stellar companion with a minimum mass of 4.6 M$_J$, in a moderately
eccentric orbit (e=0.17$\pm$0.05), 1.3 A.U. away of the star.
As the long period RV variations in red giants may also be related to a combination of effects including
stellar rotation, activity, and non-radial pulsations \citep[e.g.][]{2006A&A...457..335H}, we have analyzed the photometric data
and the behavior of line bisectors of the star, following the established practice \citep{Quelozetal2001}.

The magnitude of a photometric variability is related to a fraction of the stellar surface covered by spots. The Tycho photometry, although not very precise, gives an upper limit to this variation of 0.011 magnitude in V$_{T}$ which translates to $\sim$1$\%$ of the stellar surface covered by spots. As discussed by \citet{Hatzes2002}, with this spot coverage and the star's rotational velocity of $\sim$1 km s$^{-1}$, it is impossible to produce RV variations of the observed amplitude. In fact, their maximum amplitude would be on the order of a few m s$^{-1}$, which is comparable to the precision of our RV measurements.

\citet{Hatzes2002} has published estimates of the bisector velocity span (BVS) as a function of $v sini$ and fraction of spots. For HD 17092, we derive from their formula the BVS of a few m s$^{-1}$, which is consistent with our own BVS determination. A detailed analysis of both line bisectors and bisector velocity span shows that there is no relationship between line profile variability and the observed RV changes. In addition, line profile variations, measured with the same spectra as the ones used to obtain RVs, show no correlation with RV. 

A lack of both line profile variations and photometric variability eliminates nonradial pulsations and surface activity related to stellar rotation as a possible cause of the observed RV variations in HD 17092. Another argument against a rotation-forced RV variation is the repeatability of the observed RV changes and their strictly periodic character. Spots, as we understand them, appear and disappear on the stellar surface and the number of spots, as well as their location, vary from one cycle to another.  Therefore, a resulting variability is unlikely to maintain the same pattern over many cycles. In the case of HD 17092 the observed 3 cycles of the 360-day period are consistently fitted with a single Keplerian orbit. Consequently, the most plausible explanation of our data is the presence of a sub-stellar mass companion around the star.

A sufficiently large number of planet detections by high precision RV surveys of GK-giants 
will make them efficient tools with which to
study planet formation around intermediate-mass stars and 
the dynamical evolution of planets induced by a post-MS evolution of their parent
stars. Current constraints on a stellar mass dependence of the disk mass and the timescale
of depletion of its gas and dust components come from studies of disks around young stars.
For example, in addition to the previous work \citep[e.g.][]{2001ApJ...553L.153H}, recent Spitzer 
observations \citep[e.g.][]{2006ApJ...651L..49C}  appear to confirm
that disks around intermediate and high-mass stars have lifetimes
significantly shorter than 5 Myr. These results have direct consequences for the competing
theories of giant planet formation, because the core accumulation scenarios \citep{1996Icar..124...62P, W2000, 2004A&A...417L..25A} 
require at least a few million years for a core to
form, whereas planet formation from a disk instability \citep{1997Sci...276.1836B, 2002Sci...298.1756M} can be very short. Clearly, the searches for planets around giant
stars have a unique capability to provide the statistics which are needed to decisively constrain
the efficiency of planet formation as a function of stellar mass and chemical composition. 

It is quite reasonable to expect that the giant star surveys will have the potential to
verify predictions of the post-MS orbital evolution which emerge from numerical
simulations \citep[e.g.][]{1996Sci...274..954R, 1998Icar..134..303D, 2002ApJ...572..556D}.
In principle, each detection of a planet around a giant represents a snapshot
of the dynamical evolution of orbits around a particular, evolving star. Given a sufficient
number of planet detections it should be possible to constrain the principal evolutionary
scenarios and obtain a consistent picture of dynamical changes in planetary systems
in response to the evolution of their parent stars. An excellent example of the analysis
of planetary orbits, which would be applicable in this context has been presented by
\citet{2005Natur.434..873F}. 

HD 17092b is the tenth
published discovery of a planet around a giant star. It is one of the most distant and metal-rich giants known to host a planetary system. Further discoveries of planets around
giant stars will help to develop sufficient statistics to meaningfully address the important
questions of planet formation around intermediate-mass stars and the long-term evolution and
survival of planetary systems.

\acknowledgments
We thank Dr. Shri Kulkarni for his contribution to the initial development of this project
and the HET resident astronomers and telescope operators for support. The FTS iodine
spectrum was kindly provided by Dr. Bill Cochran.
AN, AW, GN and GM were 
supported in part by the Polish Ministry of Science and Higher Education grant 1P03D 007 30.
AW also acknowledges a partial support from the NASA Astrobiology Program. 
MK was supported by NASA through grant NNG04GM62G. 
The Hobby-Eberly Telescope (HET) is a joint project of the University of Texas at Austin, the Pennsylvania State University, Stanford University, Ludwig-Maximilians-Universit\"at M\"unchen, and Georg-August-Universit\"at G\"ottingen. The HET is named in honor of its principal benefactors, William P. Hobby and Robert E. Eberly.

\appendix

\clearpage

\begin{figure}
\includegraphics[angle=-90,scale=0.8]{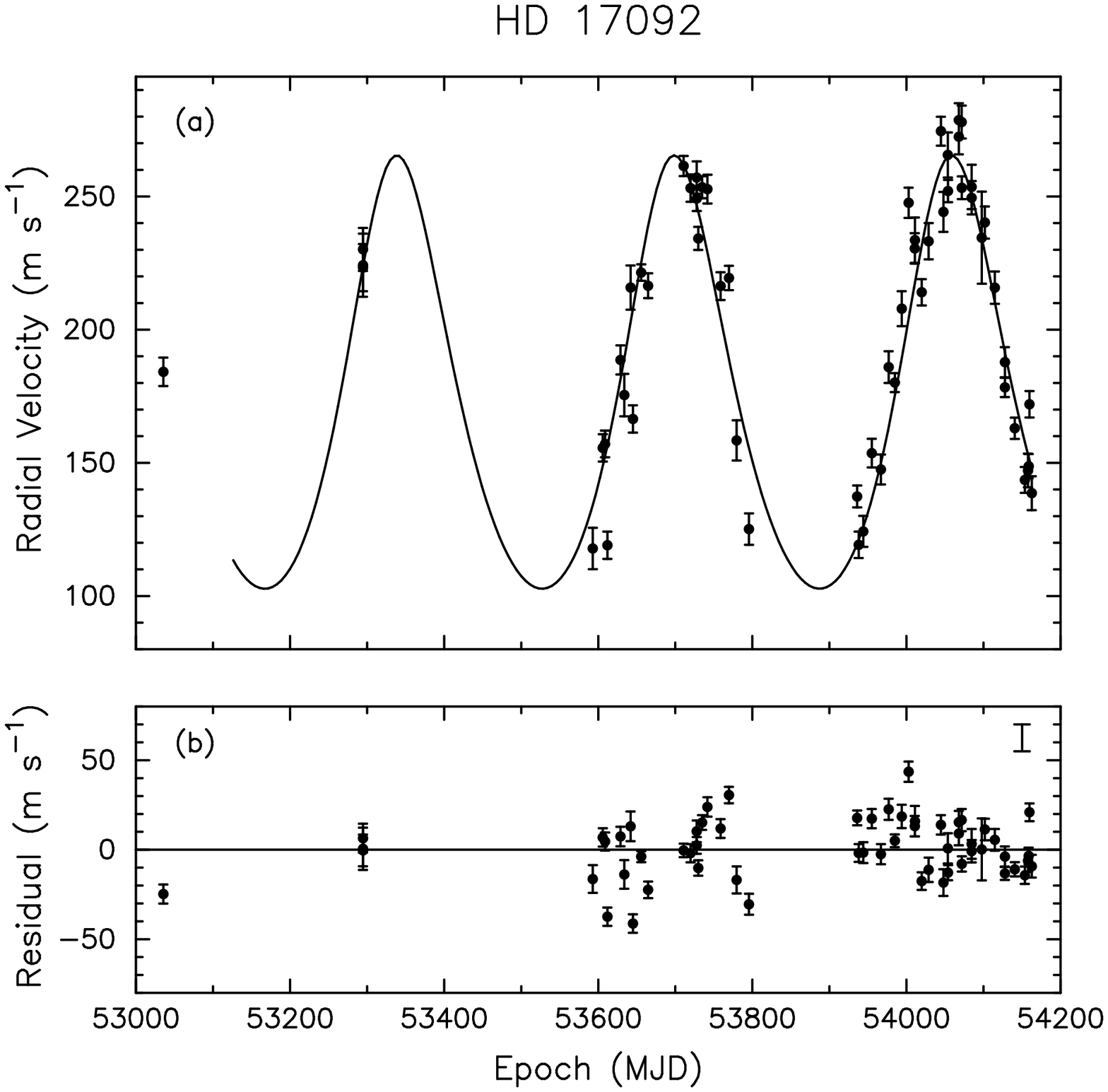}
\caption{(a) Radial velocities (filled circles) and the best-fit orbit 
(solid line) for HD 17092. 
(b) Residuals from the best-fit of a Keplerian orbit to data.
The vertical bar indicates an additional error added in
quadrature to the formal uncertainties of radial velocity measurements.}
\end{figure}

\begin{figure}
\includegraphics[angle=-90,scale=0.8]{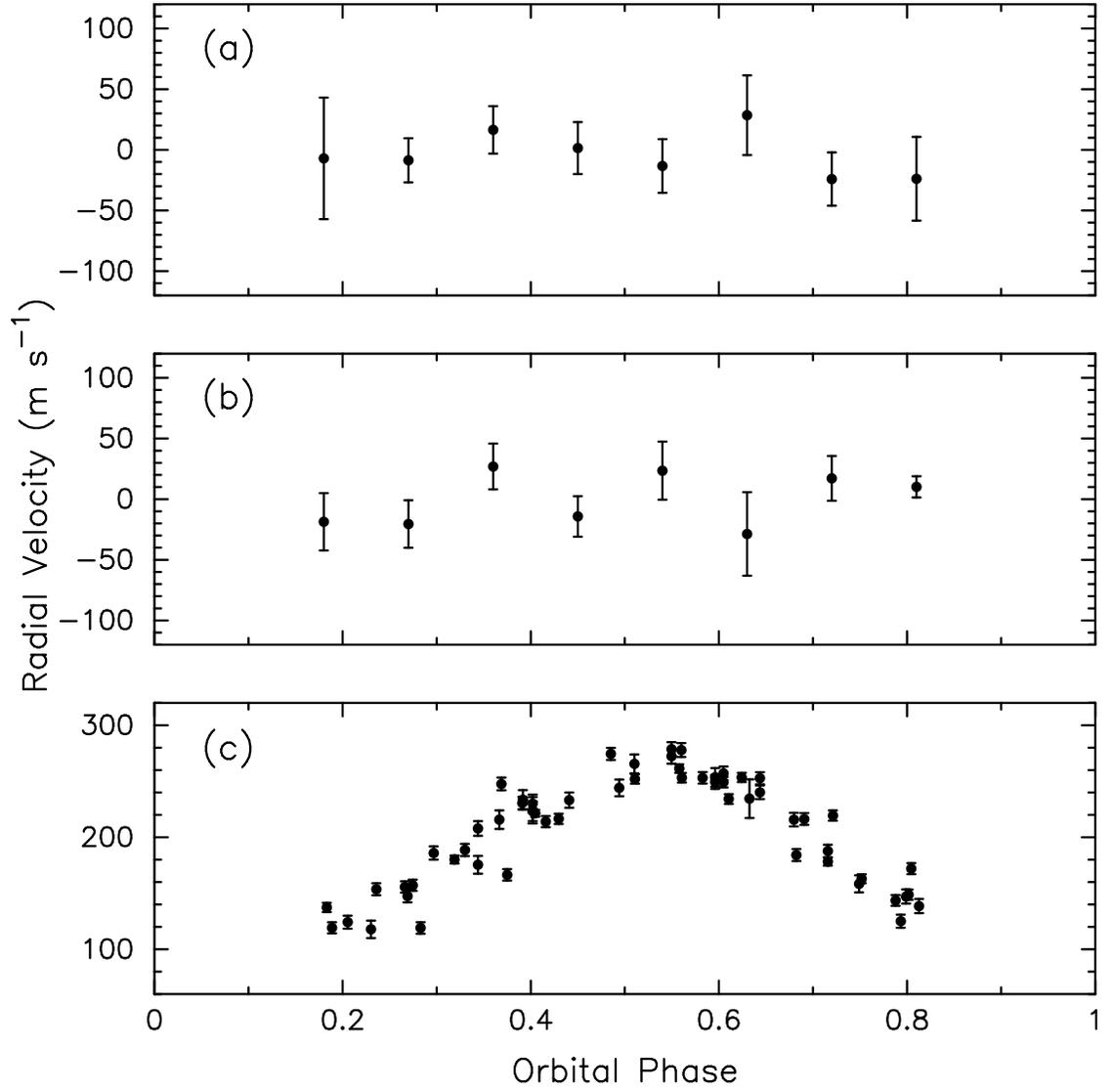}
\caption{The mean bisector velocity span (a) and curvature (b) of HD 17092,
binned as a function of orbital phase at the 0.1 interval, and compared to the measured radial
velocities of the star (c).} 
\end{figure}

\begin{figure}
\includegraphics[angle=-90,scale=0.8]{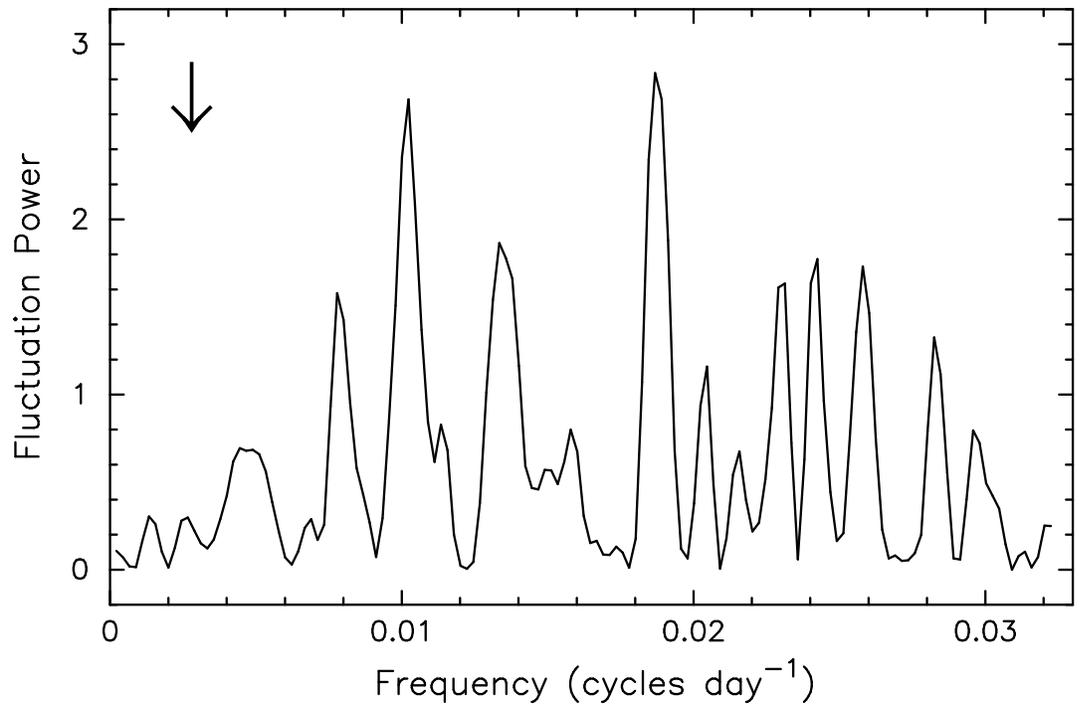}
\caption{A periodogram of the HIPPARCOS photometric measurements of HD 17092.
The vertical arrow marks the frequency corresponding to the 360-day period of the observed radial
velocity variations.}
\end{figure}

\clearpage
\begin{deluxetable}{ll}
\tabletypesize{\scriptsize}
\tablecaption{Stellar parameters of HD 17092}
\tablewidth{0pt}
\tablehead{

\colhead{Parameter} & \colhead{Value}
}
\startdata

V & 7.73\\
B-V & 1.247$\pm$0.014 \\
Spectral type & K0III \\
T${_{eff}}$ &  4650$\pm$35 \\
log g & 3.0$\pm$0.12 \\
$[$Fe$/$H$]$ & 0.18 $\pm$ 0.08\\
M$_{\star}$ & 2.3 $\pm$0.3M${_\odot}$ \\

\enddata
\end{deluxetable}
\clearpage
\begin{deluxetable}{ll}
\tabletypesize{\scriptsize}
\tablecaption{Measured and derived orbital parameters of HD 17092b}
\tablewidth{0pt}
\tablehead{
\colhead{Parameter} & \colhead{Value}
}

\startdata

P (days) & 359.9 $\pm$ 2.4 \\
T$_0$ (MJD) & 52969.5 $\pm$ 12.3 \\
K (m s$^{-1}$) & 82.4 $\pm$ 3.2 \\
a${_1}$sin i (AU) & 0.0027 $\pm$ 0.0002  \\
e & 0.166 $\pm$ 0.052 \\
$\omega$ (deg) & 347.4 $\pm$ 13.4 \\
f(m) (M$_\odot$) & 2.001$\times$10$^{-8}$ $\pm$ 1.1$\times$10$^{-9}$ \\
m${_2}$sin i (M$_J$) & 4.6 $\pm$ 0.3 \\
a$_2$ (AU) & 1.29 $\pm$ 0.05 \\

\enddata
\end{deluxetable}
\clearpage

\end{document}